\documentclass[a4paper,11pt]{article}

\usepackage{graphicx}
\usepackage{dcolumn}
\usepackage{bm}
\usepackage{comment}
\begin{document}

\newcommand{\bin}[2]{\left(\begin{array}{c} \!\!#1\!\! \\  \!\!#2\!\! \end{array}\right)}
\newcommand{\qbin}[2]{\left[\begin{array}{c} \!\!#1\!\! \\  \!\!#2\!\! \end{array}\right]_q}
\newcommand{\qbinu}[2]{\left[\begin{array}{c} \!\!#1\!\! \\  \!\!#2\!\! \end{array}\right]_{q=1}}

%
%
%

\huge

\begin{center}
Generating functions for canonical systems of fermions
\end{center}

\vspace{1cm}

\large

\begin{center}
Jean-Christophe Pain\footnote{jean-christophe.pain@cea.fr}, Franck Gilleron and Quentin Porcherot
\end{center}

\normalsize

\begin{center}
CEA, DAM, DIF, F-91297 Arpajon, France\\
\end{center}

\vspace{1cm}

\begin{abstract}
The method proposed by Pratt to derive recursion relations for systems of degenerate fermions [Phys. Rev. Lett. \textbf{84}, 4255 (2000)] relies on diagrammatic techniques. This efficient formalism assumes no explicit two-body interactions, makes possible the inclusion of conservation laws and requires low computational time. In this brief report, we show that such recursion relations can be obtained from generating functions, without any restriction as concerns the number of conservation laws (\emph{e.g.} total energy or angular momentum).
\end{abstract}


\section{\label{sec1} Introduction}

The statistical properties of canonical systems of fermions play a central role in the structural and radiative properties of multi-charged ions. Level densities, orbital populations, thermodynamic quantities, nuclear reaction rates and radiative opacities  are crucial for the studies of laboratory and astrophysical hot plasmas. Due to the large number of excited states at finite temperature, the calculation of the canonical partition function can become a difficult task, even for a reasonable number of particles. For instance, leaning on the fact that grand-canonical and canonical ensembles are equivalent in the thermodynamic limit, Kosov \emph{et al.} \cite{KOSOV08} formulated recently a method to compute averages in the canonical ensemble based on calculations in the grand-canonical ensemble (which are much easier). The technique presented by Pratt \cite{PRATT00a} to perform exact calculations of both canonical or microcanonical quantities, involves the counting of all possible arrangements of fermions. It is based on recursive algorithms, which can be considered as extensions to techniques used for multiparticle symmetrization.  Since the recursive method yields partition functions, it can be easily extended to the determination of any statistical (average) quantity. The canonical partition function $Z_A$, for a system of $A$
identical fermions populating a finite set of energy levels, $\epsilon_i$, at an inverse temperature $\beta$, reads:

\begin{equation}
Z_A(\beta)=\frac{1}{A!}\sum_{s_1\cdots s_A,\mathcal{P}(s)}\langle s_1\cdots s_A|e^{-\beta
H}|\mathcal{P}(s_1\cdots s_A)\rangle,
\end{equation}

where the $A$ distinguishable particles occupy the states $s_1\cdots s_A$, which are eigenstates of $H$, and where $\mathcal{P}(s)$ refers to the $A!$ permutations of $s_1\cdots s_A$. Pratt categorized the permutations diagrammatically, each diagram being expressed as a product of cycles. The sum over permutations is represented by the sum over diagrams. He extended the method to include conserved quantities. For instance, conservation of the $z$ component of angular momentum $J_z$ can be accomodated  by summing over only those diagrams which yield a fixed sum:

\begin{equation}\label{equaf}
M=\sum_{i=1}^gm_i,
\end{equation}

where $m_i$ represent the momentum projection of state $i$ (for instance, the number of states of a $j$ orbital is $g=2j+1$). This problem of the determination of states resulting from the coupling of $N$ particles can be carried out ``directly'' \cite{BREIT26,deSHALIT63}, but the approach becomes tedious for systems with a large number of particles. The problem of listing the terms arising in a complex configuration can be solved from elementary group theory \cite{CURL60,KARAYIANIS65} or by recursive techniques \cite{GILLERON09}. It was shown by Sunko and Svrtan \cite{SUNKO85,SUNKO86} and in a somewhat different way by Katriel \cite{KATRIEL83,KATRIEL89}, that recursion relations can be obtained from a set of generating functions for the number of many-body sets, with the use of Gaussian polynomials (or $q-$binomial coefficients),  In this brief report, we show that such a procedure can be extended at finite temperature without any restriction about the number of conservation laws. This enables one to recover the results of Pratt \cite{PRATT00a} and to draw a correspondence between the diagrammatic technique and the generating functions. The generating functions are introduced in Sec. \ref{sec2}, and the procedure to obtain the recursive relations is described in Sec. \ref{sec3}. Section \ref{sec4} is the conclusion.


\section{\label{sec2} From $q-$binomial coefficients to generating functions}

It was shown by Sunko and Svrtan \cite{SUNKO85} that the multiplicities of the total angular momentum projections for $n$-particle systems are given by the coefficients of some $q-$binomial coefficients \cite{ANDREWS76}. Any polynomial in $x$ may be written in terms of its roots $x_i$. By collecting powers of $x$, one gets \cite{BALANTEKIN01}:

\begin{equation}
\prod_{i=1}^g(x-x_i)=\sum_{k=0}^g(-1)^kf_k(s_1,\cdots,x_g)~x^{g-k}
\end{equation}

where $f_k$ is the $k^{th}$ elementary symmetric functions 

\begin{equation}
f_k(x_1,\cdots,x_g)=\sum_{1\leq i_1<\cdots< i_k\leq g} x_{i_1}\times\cdots\times x_{i_k},
\end{equation}

where the summation extends over all $\{i_1,\cdots,i_k\}$ such that $1\leq i_1<\cdots< i_k\leq g$. For instance,

\begin{eqnarray}
f_3(x_1,x_2,x_3,x_4)&=&x_1x_2x_3+x_1x_2x_4\nonumber\\
& &+x_1x_3x_4+x_2x_3x_4.
\end{eqnarray}

The number of states satisfying Eq. (\ref{equaf}) is equal to the coefficient of ($q^Mt$) in the polynomial \cite{KATRIEL89} 

\begin{equation}
\prod_{i=1}^g(1+q^it)=\sum_{r=0}^gt^kf_k(q,q^2,q^3,\cdots,q^g),
\end{equation}

which can be written

\begin{equation}
\prod_{i=1}^g(1+q^it)=\sum_{r=0}^gq^{r(r+1)/2}\qbin{g}{r}t^r,
\end{equation}

where $\qbin{g}{r}$ is the $q-$binomial coefficient (since we are focusing on fermions, $g$ is an even integer):

\begin{equation}
\qbin{g}{r}=\frac{(1-q^g)(1-q^{g-1})\cdots(1-q^{g-r+1})}{(1-q)(1-q^2)\cdots(1-q^r)}.
\end{equation}

The $q-$binomial coefficient is a generalization of the ordinary binomial coefficient, which corresponds to the particular case $q=1$:

\begin{equation}
\qbinu{g}{r}=\bin{g}{r}.
\end{equation}

For instance, considering $j=7/2$ and $n=3$, we have

\begin{eqnarray}
q^{-15/2}\qbin{8}{3}&=&q^{-15/2}+q^{-13/2}+2~q^{-11/2}+3~q^{-9/2}\nonumber\\
& &+4~q^{-7/2}+5~q^{-5/2}+6~q^{-3/2}+6~q^{-1/2}\nonumber\\
& &+6~q^{1/2}+6~q^{3/2}+5~q^{5/2}+4~q^{7/2}\nonumber\\
& &+3~q^{9/2}+2~q^{11/2}+q^{13/2}+q^{15/2}.
\end{eqnarray}


\section{\label{sec3} From generating functions to recursive relations}

Following Pratt \cite{PRATT00a}, let us consider a system of $A$ identical fermions populating a finite set of energy levels, $\epsilon_i$, at an inverse temperature $\beta$, and subject to the constraint $M=\sum_{i=1}^gm_i$. In that case, the partition function, noted $Z_{A,M}(\beta)$, is the coefficient of $x^M$ in the Taylor-series development of $F_A(x)$, defined as

\begin{equation}
F_A(x)=\frac{1}{A!}\frac{\partial^A F(x,z)}{\partial z^A}\Big|_{z=0},
\end{equation}

where $F(x,z)$ is the generating function

\begin{equation}
F(x,z)=\prod_{i=1}^g(1+ze^{-\beta\epsilon_i}x^{m_i}).
\end{equation}

After a first derivative, one has:

\begin{eqnarray}
F_A(x)&=&\frac{1}{A!}\sum_{i=1}^gx^{m_i}e^{-\beta\epsilon_i}\times\nonumber\\
& &\frac{\partial^{A-1}}{\partial
z^{A-1}}\prod_{k=1,k\ne i}^g(1+ze^{-\beta\epsilon_k}x^{m_k})\Big|_{z=0},
\end{eqnarray}

which is equivalent to

\begin{eqnarray}
F_A(x)&=&\frac{1}{A!}\sum_{i=1}^gx^{m_i}e^{-\beta\epsilon_i}\times\nonumber\\
& &\frac{\partial^{A-1}}{\partial z^{A-1}}\frac{\prod_{k=1}^g(1+ze^{-\beta\epsilon_k}x^{m_k})}{(1+ze^{-\beta\epsilon_i}x^{m_i})}\Big|_{z=0}.
\end{eqnarray}

Using Leibniz formula for the multiple derivative of a product of two functions, one obtains:

\begin{eqnarray}
F_A(x)&=&\frac{1}{A!}\sum_{i=1}^gx^{m_i}e^{-\beta\epsilon_i}\times\nonumber\\
& &\sum_{n=0}^{A-1}\bin{A-1}{n}\left[\frac{\partial^{A-1-n}}{\partial z^{A-1-n}}\frac{1}{(1+ze^{-\beta\epsilon_i}x^{m_i})}\right]\times\nonumber\\
& &\prod_{k=1}^g(1+ze^{-\beta\epsilon_k}x^{m_k})\Big|_{z=0}.
\end{eqnarray}

Since

\begin{equation}
\frac{\partial^k}{\partial z^k}\frac{1}{(1+az)}=\frac{k!(-1)^ka^k}{(1+az)^{k+1}},
\end{equation}

one finds:

\begin{equation}
F_A(x)=\frac{1}{A}\sum_{n=1}^A(-1)^{n-1}e^{-n\beta\epsilon_i}\left[\sum_{i=1}^gx^{n.m_i}\right]F_{A-n}(x),
\end{equation}

which leads, since $Z_{A,M}(\beta)$ is the coefficient of $x^M$, to the following relation

\begin{equation}
Z_{A,M}(\beta)=\frac{1}{A}\sum_{n=1}^A\sum_{i=1}^ge^{-n\beta\epsilon_i}(-1)^{n-1}Z_{A-n,M-nm_i}(\beta),
\end{equation}

which is exactly Eq. (6) of Ref. \cite{PRATT00a}. Therefore, the number of states with total energy $E$ is obtained by ($T\rightarrow\infty$, \emph{i.e.} $\beta\rightarrow 0$):

\begin{equation}
N_{A,E,M}=\frac{1}{A}\sum_{n=1}^A\sum_{i=1}^g(-1)^{n-1}N_{A-n,E-n\epsilon_i,M-nm_i}.
\end{equation}

Such a formalism can be extended to include additional constraints. For instance, in LS coupling, the total number of states with total energy $E$, reads :

\begin{eqnarray}
N_{A,E,M_L,M_S}&=&\frac{1}{A}\sum_{n=1}^A\sum_{i=1}^g(-1)^{n-1}\times\nonumber\\
& &N_{A-n,E-n\epsilon_i,M_L-nm_{\ell i},M_S-nm_{s i}}.
\end{eqnarray}

Such recursion relations for the angular-momentum states are of great interest in detailed line-by-line calculations of atomic-structure and radiative opacity. For instance, they enable one to determine the size of the blocks of the hamiltonian matrix corresponding to a given value of the total angular momentum $J$, and to obtain the number of lines of a transition array. They can also play a role in the calculation of the rates of processes such as autoionization. Applications related to atomic spectroscopy will be presented in a future paper.

In the present work, as in Ref. \cite{PRATT00a}, no explicit two-body interactions are considered, \emph{i.e.} it is assumed that the total energy of the system is a linear function of the numbers of particles. However, the interactions could be included, in an approximate way, either in a perturbative manner as suggested by Pratt \cite{PRATT00a}, or following the variational Jensen-Feynman approach \cite{PAIN09}, which consists in optimizing the one-particle energies $\epsilon_i$ in order to obtain the best agreement with the exact free energy. Such an approach is driven by the ability to obtain robust recursion relations \cite{GILLERON04,WILSON07} for the calculation of partition functions and introduces no further difficulties in the averaging process. 


\section{\label{sec4} Conclusion}

In this report, it was shown that the generating function method, formulated for instance through $q-$binomial coefficients, and which was introduced for the numbering of angular-momentum states of an arbitrary configuration, can be easily extended to calculate any average quantity in the canonical ensemble. The technique enables one to recover the recursive relations for the partition functions obtained by Pratt using diagrammatic techniques.

\end{document}